\documentclass[preprintnumbers,amsmath,amssymb]{revtex4}

\usepackage{graphicx} 
\usepackage{times}
\usepackage{amsmath}
\usepackage{amssymb}
\usepackage{units}

\usepackage{dcolumn}
\usepackage{bm}

\begin{document}

\title{ANTARES, a scanning photoemission microscopy beamline at SOLEIL}

\author{Jos\'{e} Avila, Ivy Razado, Stehane Lorcy, Bruno Lagarde, Jean-Luc Giorgetta, Fran\c{c}ois Polack and Mar\'{i}a C. Asensio}

\address{Synchrotron SOLEIL, L'Orme des Merisiers, Saint Aubin-BP 48, 91192 Gif sur Yvette Cedex, France}

\date{\today}

\begin{abstract}

As one of the latest beamline built at the SOLEIL synchrotron source, ANTARES beamline offers a spectroscopic non-destructive nano-probe to study advanced materials. This innovative scanning  photoemission microscopy combines linear and angle sweeps to perform precise electronic band structure determination by Nano Angle Resolved Photoelectron Spectroscopy (Nano\textendash ARPES) and chemical imaging by core level detection. The beamline integrates effectively insertion devices and a high transmission beamline optics. This photon source has been combined with an advanced microscope, which has precise sample handling abilities. Moreover, it is fully compatible with a high resolution R4000 Scienta hemispherical analyzer and a set of Fresnel Zone Plates (FZP) able to focalize the beam spot up to a few tenths of nanometers, depending on the spatial resolution of the selected FZP. We present here the main conceptual design of the  beamline and endstation, together with some of the firsts commissioning results. 

\end{abstract}

\maketitle
\section{Introduction}

The use of synchrotron radiation is important for many applications in material science, specially for soft x-ray photoemission spectroscopic experiments, where photoelectrons emitted by the investigated samples are measured as a function of their kinetic energy when constant photon energy is used.  Electron and positron storage rings together with insertion devices are essential for creating brilliant x-ray beams in a wide spectral range. Owing to the increasing demand for improved lateral resolution determinations, additional microscopy-dedicated synchrotron radiation beamlines are under construction worldwide,  with high brilliant and intense sources of soft x-rays.

The new ANTARES beamline is installed at one of the long straight-line of the SOLEIL storage ring in France [1]. This source usually operates at an electron energy of 2.5 GeV, injection currents of 500 mA and \textquotedblleft Top-up$\textquotedblright$ mode.  The beamline was designed for  Nano-ARPES measurements, core level imaging and x-ray absorption spectroscopy in the energy range between $\sim$ 10 and 900 eV, making use of the intense radiation emitted by two undulators installed in a tandem configuration. Linear and circular polarized x-rays are available by a fast and easy exchangeable procedure. Accordingly, a plane-grating monochromator (PGM) is mounted with unique capabilities combining a good suppression of higher orders light with a superior resolving power energy, affecting only moderately the high photon transmission of the beamline. This monochromator is an innovative simplified Petersen's scheme [2], which is characterized by a slitless entrance and the use of two Varied Linear Spacing (VLS) gratings with a Variable Groove Depth (VGD) along the grating lines. The use of VGD, particularly, improves the higher order light suppression. In fact, the focussing function accomplished by the sagittal mirror is replaced here by the VLS grating, with holographic focussing effect. As a result, the monochromator is constituted only by two plane VLS with VGD gratings, and a plane mirror with a fixed exit slit. 

The whole special design of the ANTARES beamline optics ensures an homogeneous and a coherent illumination of  a set of Fresnel Zone plates (FZP) able to cover a wide photon energy domain. Up to now, there  are several beamlines that offer the same soft x-ray spectral range and high lateral spatial resolution (i.e. Photoemission electron microscopy or PEEM beamlines) ; however, energy and momentum resolution of the obtained data are typically bellow 200 meV and 0.05 $\AA$$^{-1}$, respectively. These figures of merit are hardly compatible with the precision required by a typical electronic structure determination using conventional ARPES measurements.

\begin{figure}[h]
\includegraphics[width=38pc]{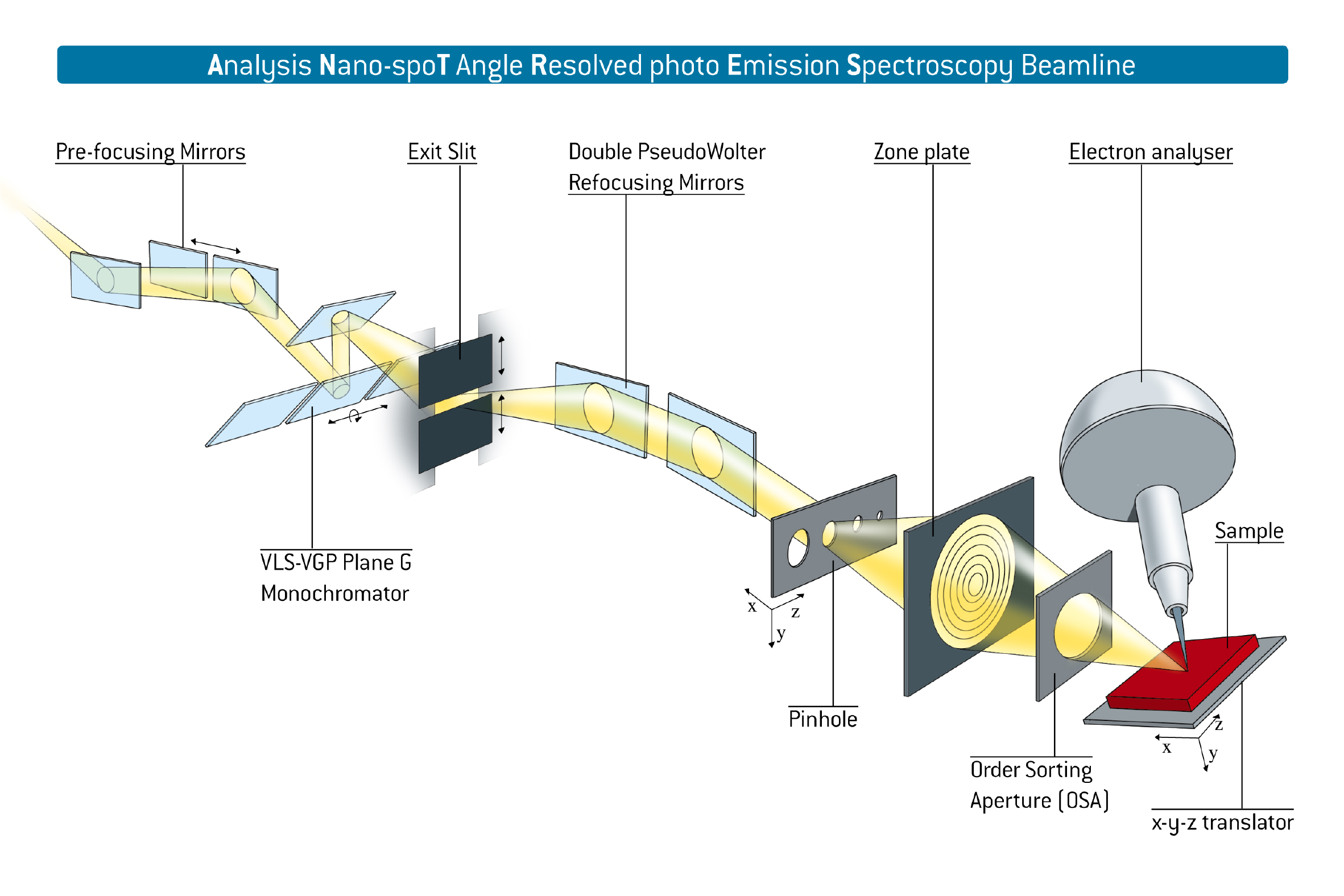}\hspace{2pc}%
\begin{minipage}[b]{38pc}\caption{\label{label}Schematic representation of the basic components of the ANTARES beamline. The first pre-focusing mirrors collimate the undulators beam to a plane grating monochromator with two gratings that can be exchanged by remote control. The second set of mirrors or pseudo Wolter post-localization mirrors refocuses the beam alternatively onto the sample or the pinhole, depending if a micro- or a nano-meter spot size is required.  The FZP focusing system is also included in the Nano-ARPES endstation, together with an order selection aperture (OSA) placed between the FZP and the sample, in order to suppress unwanted diffraction orders.}
\end{minipage}
\end{figure}

Since the ANTARES beamline combines the use of nano-spot x-ray illumination, precise sample scanning and high resolution state-of-the-art hemispherical analyzer, very interesting important fields of research can be addressed. In the succeeding sections, we describe the optical layout and the performance of the beamline including several experimental results recently obtained.
 
\section{X-ray source and beamline }

The design of the beamline optics satisfies the demands imposed by a high-energy resolution performance together with a high-flux of polarized photons spanning a wide energy range, from 10 to 1000 eV. The beamline includes two insertion devices with a high transmission optics (see Fig. 1), whose main components are: a pre-focalisation optics, a PGM and a Pseudo-Wolter as a post-focalisation optics. The high stability of the re-focalisation system composed by two alternative sets of spheric and toric mirrors in horizontal deviation, allows working alternatively with micro- or nano-spot light. One couple of mirrors is able to illuminate directly the specimen while the second set of mirrors illuminates the sample passing first through a pinhole and a FZP.  The PGM, equipped with a plane mirror and two VLS, is only focused by the holographic effect of the gratings. Each grating, produced by holographic etching  (Horiba-Jobin-Yvon) is extensively exploited throughout almost the full range of the beamline energies thanks to a lateral grading of the groove depths.  Taking advantage of the lamellar gratings with VGD, ANTARES beamline combines a high resolving power, an excellent harmonic rejection ratio and a high flux throughout the working energy range, which is particularly compatible with the extremely demanding use of FZPs. Typically, the photon flux on sample is 6.4 x 10$^{\ 12 }$ photons/sec/0.01\% BW working in micro-spot mode and 5.4 x 10$^{\ 10 }$ photons/sec/0.01\% BW for nano-spot experiments, with a standard resolving power of 32000 at 200 eV photon energy. 

\section{Experimental endstation and Nano-ARPES microscope}

The experimental end-station, is equipped with an analysis chamber containing the microscope, a preparation chamber with gas lines and a fast entry chamber, that allows an effective and fast sample transfer from atmospheric to ultra high vacuum conditions.   The hutch accommodates the pinhole chamber and the analysis chamber that contains  the electron analyzer and nano-spot focusing optics. In order to ensure a precise nanometer scanning of the sample both the thermal drift and the mechanical vibrations are minimized by an interferometric control. The thermal variation is stabilized to be lower than 0.1$^{\circ }$C and the mechanical vibrations are minimized to have only spurious displacements not bigger than  5 nm. The main difference of the ANTARES microscope from other conventional ARPES instruments is that the specimens can be mounted on a high-precision plate that ensures their nanoscale positioning in the x, y and z directions. The polar angle ($\Theta$) and the azimuthal angle ($\Psi$) can also be automatically scanned over a 90$^{\circ }$ range. Finally, the soft x-ray beam with a controlled linear or circular polarization can be focused to about 30 nm (or better), using FZP lenses. The ANTARES microscope has two operating modes, spectroscopy with nano-spot and spectroscopic imaging.

\begin{figure}[h]
\includegraphics[width=28pc]{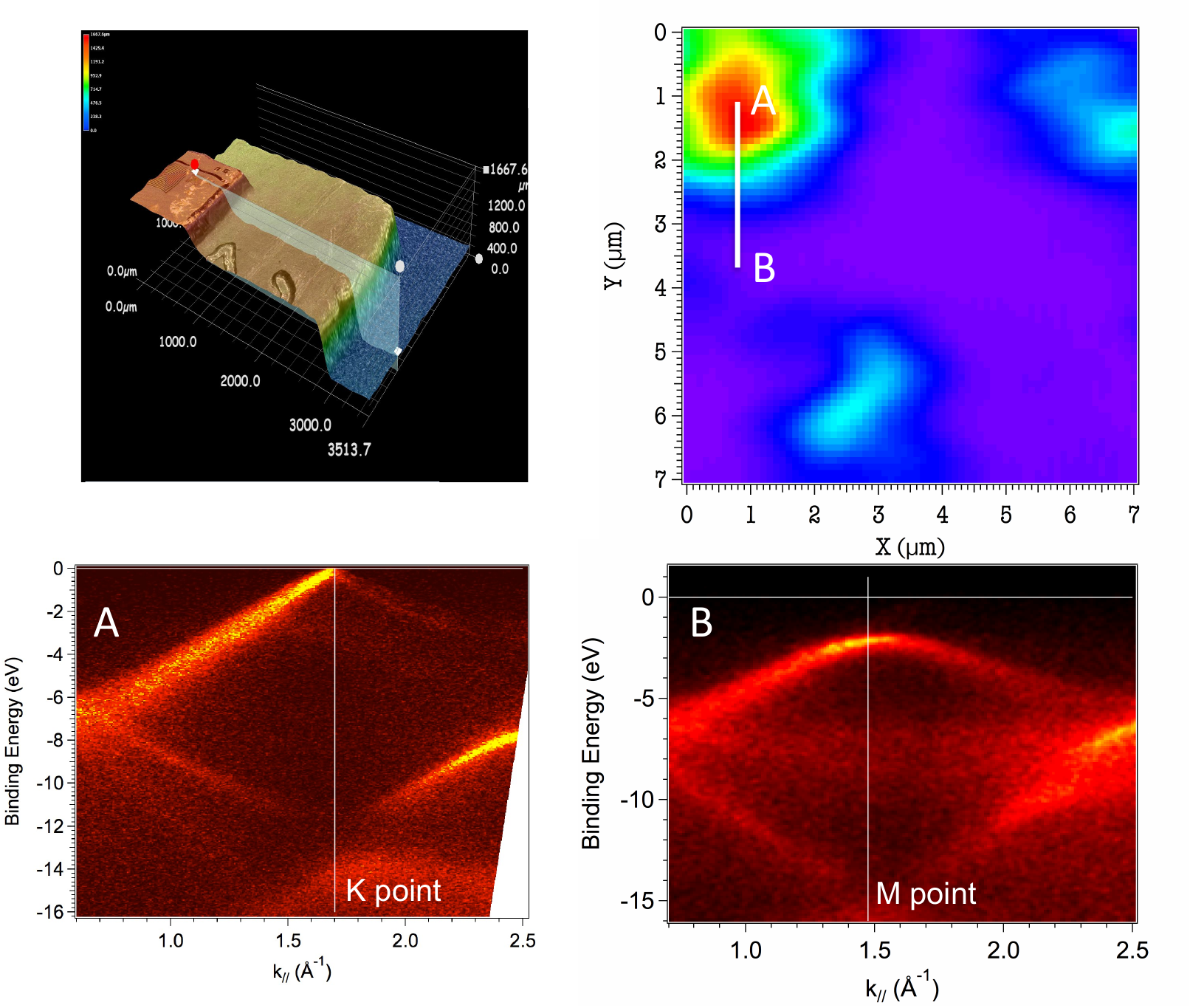}\hspace{2pc}\centering
\begin{minipage}[b]{38pc}\caption{\label{label} Top left panel shows the visualization of a HOPG by optical microscopy. Spatially and angular (bottom panels)resolved photoelectron intensity maps of a HOPG,  top right and bottom panels, respectively.}
\end{minipage}
\end{figure}
  
\section{Nano-ARPES microscope experiments}

To demonstrate the capabilities of the new microscope, we present the first results obtained from a specimen of highly oriented pyrolytic graphite (HOPG). The sample is a poly-crystal composed of grains of micrometer-sized single-crystals of graphite randomly oriented in the basal plane of the crystal. The basic unit of the HOPG is graphite, which has a planar structure, wherein each layer, carbon atoms are arranged in a hexagonal lattice. The sp$^{2}$ electrons of the carbon atoms in each plane are bonded by strong covalent  (sigma) bonds and covalent (pi) bonds for their other p electrons.  Figure 2 shows a 7 x 7$\mu$m map of the photoemission intensity of a reduced energy window around the Fermi level, which is the energy that separates the occupied from the unoccupied bands. The microscopic detection geometry was fixed so that it can only detect the intensity from those grains oriented in the $\Gamma$K direction, coincident with the energy window of the detector. The thermal and mechanical stability of the microscope makes it possible to obtain high contrast and high reproducibility of the acquired images. The visualization of the different grains, each measuring 1-2 microns, is direct and fast. Moreover, the Nano-spot mode of the microscope allows performing a full spectroscopic and electronic band dispersion determination not only inside the grain but also in its surroundings.

Thus, the new Nano-ARPES microscope recently installed on ANTARES, is already capable of providing spectroscopic images with a spatial resolution of several tens of nanometers, while preserving angular and energy resolutions comparable to the best performing ARPES instruments installed nowadays in synchrotron radiation sources.

\section{Acknowledgments}
The authors acknowledge the valuable support of the Engineering service of the Synchrotron SOLEIL, in particular for the design of the Nano-ARPES microscope and  the optical definition of the ANTARES beamline. The Synchrotron SOLEIL is supported by the  CNRS and the CEA, France.

\smallskip
 
\end{document}